\documentclass[12pt]{article}

\usepackage{graphicx}
\usepackage{amssymb,amsmath}
\usepackage{url}
\usepackage{natbib}
\usepackage[format=hang,font=small,labelfont=bf]{caption} 
\usepackage[none]{hyphenat} 
\usepackage{hyperref}

\bibpunct{(}{)}{;}{a}{}{,} 

\title{Enhanced Learning with Web-Assisted Education}

\author{Anna Helga Jonsdottir\thanks{PhD student, University of Iceland, Dunhaga 5, 107 Reykjavik, Iceland} \and  Gunnar Stefansson\thanks{Professor, University of Iceland, Dunhaga 5, 107 Reykjavik, Iceland}}
\date{}
\begin{document}

\maketitle

\subsection*{Keywords}
web-based education; statistics education; randomized crossover experiment; item response theory

\section*{Abstract}
An educational system, the tutor-web (\url{http://tutor-web.net}), has been developed and used for educational research.
The system is accessible and free to use for anyone having access to the Web. It is based on open source software and the teaching material is licensed under the Creative Commons Attribution-ShareAlike License. The system has been used for computer-assisted education in statistics and mathematics. It offers a unique way to structure and link together teaching material and includes interactive quizzes with the primary purpose of increasing learning rather than mere evaluation.

The system was used in a course on basic statistics in 2011. Three types of data were gathered during the course. A randomized crossover experiment was conducted to assess the possible difference in learning (measured by repeated exams) between students using the system and students doing regular homework. The difference between the groups was not found to be significant. Responses to quiz questions were collected and analysed with item response theory type models. 
These analysis were used to improve the item banks.
Finally, the students answered an in-class survey regarding their experience using the tutor-web. The responses of the students gave clear indications of student preferences.

\section{Introduction}
The tutor-web is a web-based educational system accessible and free to use for everyone having access to the Web.
The teaching material in the system is linked together in a unique way making it easy for the student to browse through the material. The system also includes interactive quizzes with the primary purpose of increasing learning rather than mere evaluation.
The system is solely based on open source software. It is written in Plone \citep{nagle2010plone} which is a Web-based content management system (CMS) 
built on top of the Zope Application Server \citep{latteier2002zope}. The teaching material is licensed under the Creative Commons Attribution-ShareAlike License \footnote{http://creativecommons.org/} to provide usage of material to institutions of limited resources. 
The tutor-web has been used in several courses at the University of Iceland. In the following analysis, data collected in a basic statistical course taught in 2011 will be used. The goal of the analysis is to answer three research questions:
\begin{enumerate}
 \item Is there any difference in learning between students using the tutor-web and students doing traditional homework?
 \item Are the item banks (collection of quiz questions) adequate?
 \item What do the students think about the tutor-web?
\end{enumerate}
It is of interest to measure possible difference in learning between students answering quiz questions in the tutor-web and students doing more traditional homework such as handing in written assignments. While no additional work is needed by teachers after tutor-web assignments have been handed in, written solutions needs to be corrected which can be very time consuming and therefore costly. The system could therefore save time and effort if it can replace traditional homework to some extent. To answer this question a randomized trial was conducted where students either worked within in the tutor-web as homework or handed in written assignments corrected by a teaching assistant. 

It is important that the item banks/item pools (collection of quiz questions) used in the quizzes consist of items that fit students with different ability levels, that is, they need to have easy items up to very difficult ones. In order to investigate whether the item banks used in the experiment fulfill this requirement models within the item response theory (IRT) framework \citep{lord1980applications} were fitted to the students' responses to the quiz questions.

In order to measure the students satisfaction with the system they were asked to answer some questions regarding their experience using the system at the end of the course.

A short introduction to the tutor-web will be given in Section \ref{tw}. The data used in the analysis and the models applied are described in Section \ref{methanddata} followed by some results and discussion in Section \ref{results}. Sections \ref{methanddata} and \ref{results} are split up to three subsections each dealing with one research question. Future work can be found in Section \ref{futurework} and summary of conclusions in Section \ref{conclusion}.

\section{The tutor-web \label{tw}}
A short introduction to the tutor-web is given in the following section. For a more detailed description see \cite{stefansson2004tutor}.
\subsection{The organization of educational material}
\begin{figure}[t]
\caption{The structure of the tutor-web.}
\centering\includegraphics[scale = 0.4,trim=0mm 55mm 0mm 0mm]{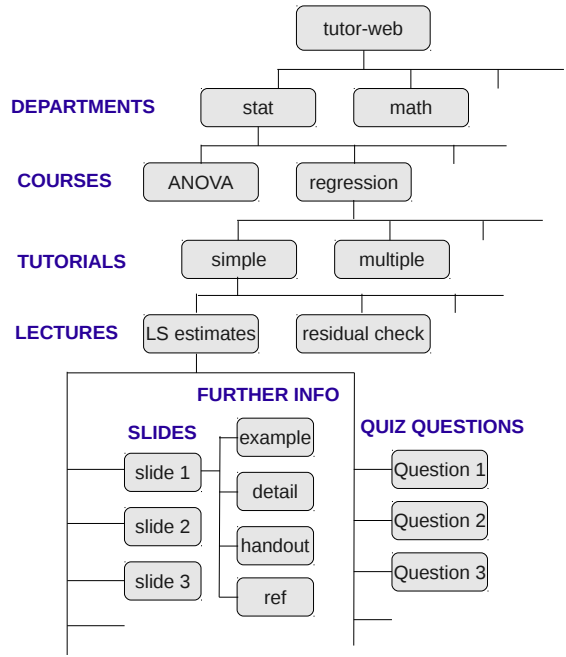}
\label{structure}
\end{figure}

The teaching material in the system is organized into a tree (Fig. \ref{structure}). Slides are grouped together to form lectures which are grouped into tutorials. A tutorial can belong to more than one course and should be built up around a single theme. For example, a tutorial on simple linear regression could both be a part of a general course on regression and an introductory statistics course. Quiz questions are linked to every lecture (see Section \ref{qquestions}). 

Four types of users are defined in the tutor-web: regular users, students, teachers and managers.
All tutor-web users can view and download teaching material but in order to answer quiz questions the user needs to become a tutor-web student by filling out a simple form. The student also needs to agree to that grades he or she earns in the system are recorded into a database and maybe used anonymously for research purposes. A tutor-web teacher can add and edit tutorials, lectures, slides and quizzes while managers can in addition add and edit departments and courses and give teacher rights. 

\subsection{Adding and viewing material}
Teaching material can be added easily to the system through a web browser. A tutor-web teacher can built up a new tutorial with lectures including collection of slides. The slide is the core unit of the tutor-web. The other units are simply collection of information from the slides. A slide has a title, some text and/or figure(s). The format of the text can be LaTeX \citep{goossens1994latex}, plain text or HTML. The figure(s) can be uploaded files (png, gif or jpeg) or they can be rendered from a text based image format (R \citep{R2011} or Gnuplot \citep{williamsgnuplot}). The teacher can choose to link some additional material to a slide such as examples, more details and/or references. 

\begin{figure}[t]
\caption{Different views into the database of teaching material in the tutor-web.}
\centering\includegraphics[scale = 0.27]{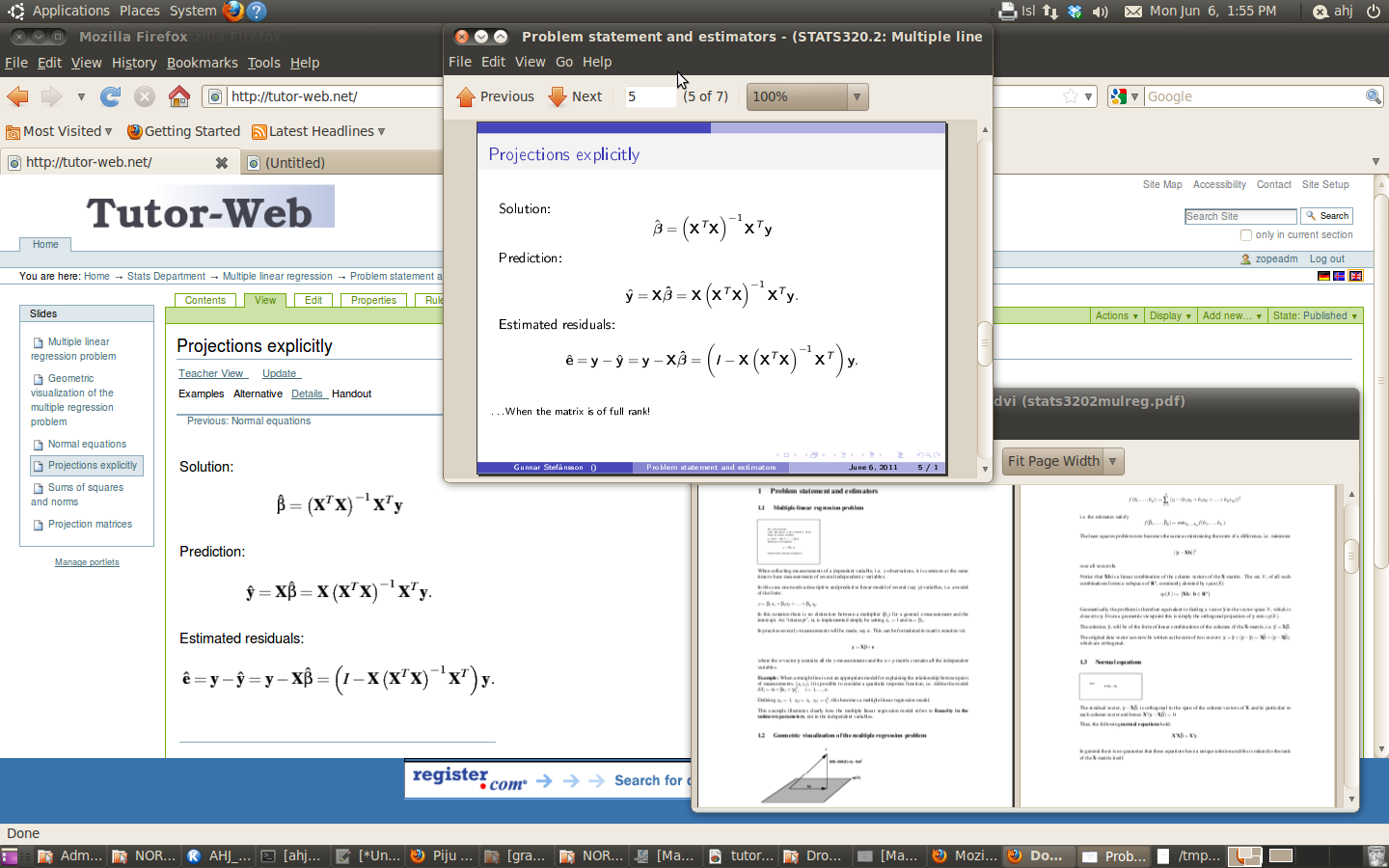}
\label{views}
\end{figure}

There are three different ways for a tutor-web user to view the teaching material (Fig. \ref{views}). The user can enter a lecture and browse through the material slide by slide. Links are provided to additional material attached to the slide (examples, references, ...) if any. The user can also download a PDF document including all the slides belonging to the lecture. The slides do not include the additional material linked to them (if any). The slides, in PDF format, are made with the LaTeX package Beamer \citep{number2011beamer} and should be ready to be used in a classroom. The third way of viewing teaching material in the tutor-web is on the tutorial level. Users can download a PDF document including all lectures belonging to that tutorial. A figure of the slides is provided in the document along with all additional material attached to them providing a handout including all relevant information. In a fully developed tutorial this corresponds to a complete textbook.

\subsection{Quiz questions and grading \label{qquestions}}
Quiz questions (items) are grouped together so they correspond to material within a lecture. The primary purpose of the tutor-web quizzes is to increase learning rather than mere evaluation. Students should therefore be allowed to answer quizzes at their own leisure. There is no built-in penalty for late answers, but an instructor may of course decide to specify a date at which a grade is extracted from the system. 

Questions and answers can be added to the system through a browser or be uploaded from a file. A quiz question can have as many answers as desired and there is an option to randomize the order of the answers. The format of the text can be LaTeX, plain text or Structured text. The system also allows the use of the statistical package R \citep{R2011} when a question is generated. This allows the generation of similar but not identical data sets and graphs for students to analyse or interpret. 

Currently a student gets one point for answering a question correctly and -1/2 for a wrong answer. Since the purpose of the quizzes is to allow the students to learn and thus improve the grade, only the last eight questions the student answers are used when the grade is calculated for each lecture. The student can track the grade with a press of a button, allowing each individual to monitor personal progress.

When data used in this paper was collected the question given to a student answering a quiz was selected randomly with uniform probability.
This has since been changed such that when the student's grade goes up the questions become more difficult (Section \ref{futurework}).

\section{Methodology and data \label{methanddata}}
The tutor-web has been used in several courses taught at the University of Iceland. The data analysed here was gathered in a basic statistics course in 2011. The teaching material and item banks used in the course were written in Icelandic but an English version can be found by entering the tutor-web (\url{http://tutor-web.net}), chose \texttt{Stats department} and from there  \texttt{Introductory statistics} (a direct link: \url{http://www.tutor-web.net/tutor-web/stats/stats201}).
The students attending the course came from two different programs at the university and had very different mathematical background. A bit over half of the students  had taken a course in calculus the year before and had strong mathematical background while the rest had much weaker background, many had not done any math for years. 

In order to answer the three research questions three types of data were gathered and three different analysis performed. 

\subsection{Analysis of difference in learning}
\begin{figure}[t]
\caption{The design of the experiment.}
\centering\includegraphics[scale = 0.45]{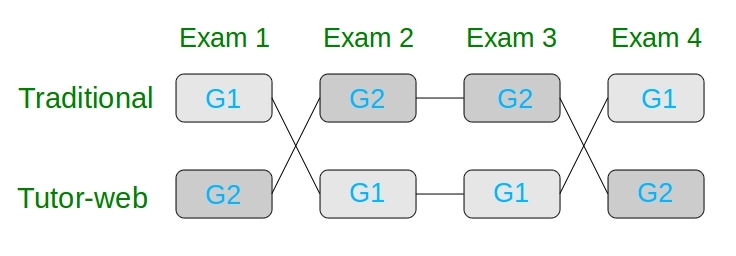}
\label{design}
\end{figure}

An experiment was conducted to assess potential difference in learning between students using the tutor-web and students doing regular homework. The experiment was a randomized crossover experiment (Fig. \ref{design}. ). The students were randomly split into two groups. In the first part of the experiment half of the students worked on quizzes in the tutor-web as homework while the other half handed in a written assignment. Shortly after the students handed in their homework they took an unexpected exam in class. In the next part of the experiment the groups were crossed, that is, the students that worked in the tutor-web before were told to hand in a written assignment and vice versa and again the students were tested. This was repeated two more times with 157 students taking at least one exam. The four subjects covered in the assignments and exams were discrete distributions, continuous distributions, hypotheses concerning means and contingency tables. 

In order to test if there was any difference in learning in the groups, it was decided to use a mixed effect model with the exam scores as the response variable:
\begin{equation}
y_{ijkl} = \beta_{1i} + \beta_{2j} + \beta_{3ij} + \beta_{4k} + b_{l} + \epsilon_{ijkl}
\label{anova}
\end{equation}
where $\beta_1$ is the fixed treatment effect, traditional/tutor-web ($i = 1,2$), $\beta_2$ is the fixed math experience effect ($j = 1,2$), $\beta_3$ is the interaction between the treatment effect and the math experience effect, $\beta_4$ is the exam effect ($k = 1,2,3,4$) and $b$ is the random student effect ($l = 1,2,...,157$). The full model was fitted to the data and insignificant terms removed.
\subsection{Analysis of quiz question responses}
While the students took part in the experiment described above quiz question responses were gathered. It is important that the question banks include questions with a broad range of difficulty levels. In order to investigate whether that is the case for the four banks of questions used, models from item response theory (IRT) were fitted to the data. IRT is the framework used in psychometrics for the design, analysis, and grading of computerized tests to measure abilities \citep{lord1980applications}. Even though the primary purpose of the analysis here is not to grade the students the models can be used to estimate the difficulty level of the items. The models give the probability of answering an item correctly given some characteristic of the item and the ability of the student. 

The data gathered here is somewhat different from data usually used within the IRT framework where a group of people is gathered to answer every item in a bank once.
In this study items were allocated randomly so the students were exposed to some of the items more than once and other items not at all. 

The simplest model is the one parameter logistic model where only the difficulty level of the items is considered:
\begin{equation}
P(x_{im} = 1|z_m;\theta) = \frac{\exp(z_m - \beta_i)}{1 + \exp(z_m - \beta_i)} 
\label{m1}
\end{equation}
where $x_{im}$ is the response of the $m$-th student to the the $i$-th question, $z_m$ is the ability of the $m$-th student (often assumed to follow a standard normal distribution) and $\beta_i$ is the difficulty parameter of the $i$-th question. The difficulty parameter sets the location of the curve. It shifts the curve from left to right as the item becomes more difficult. Items with a negative difficulty parameter are therefore easier than items with positive difficulty parameter.  

A bit more complicated is a model where a common discrimination parameter, $\alpha$, is also estimated: 
\begin{equation}
P(x_{im} = 1|z_m;\theta) = \frac{\exp\{\alpha(z_m - \beta_i)\}}{1 + \exp\{\alpha(z_m - \beta_i)\}} 
\label{m2}
\end{equation}

The two parameter logistic model is an extension of the one parameter model where a discriminant parameter is estimated for every item:
\begin{equation}
P(x_{im} = 1|z_m;\theta) = \frac{\exp\{\alpha_i(z_m - \beta_i)\}}{1 + \exp\{\alpha_i(z_m - \beta_i)\}} 
\label{m3}
\end{equation}
The discrimination parameter is a measure on how effective an item is at discriminating between students at different ability levels.  
An item with high discriminant parameter is better at discriminating than items with low discriminant parameter.

Finally, the three parameter logistic model also includes a guessing parameter $c_i$: 
\begin{equation}
P(x_{im} = 1|z_m;\theta) = c_i + (1 - c_i)\frac{\exp\{\alpha_i(z_m - \beta_1)\}}{1 + \exp\{\alpha_i(z_m - \beta_1)\}} 
\label{m4}
\end{equation}
The guessing parameter measures how likely it is to obtain the correct answer by guessing. 

No learning is taken into account in the models above. It was therefore decided to disregard answers to questions the students had seen before in the estimation of the models. The four models were fitted to the question response data and compared using a likelihood ratio test. 

\subsection{Students feedback}
At the end of the semester the students were asked to answer some questions about their experience using the tutor-web. 121 students took part in the survey. The questions were:
\begin{enumerate}
 \item I learn by answering quiz-questions in the tutor-web (agree/disagree)
 \item I find the first questions easier than the last ones (agree/disagree)
 \item I guess the answer instead of trying to solve the problem (agree/disagree)
 \item As homework I prefer (traditional assignments/the tutor-web/mix of traditional and the tutor-web)
\end{enumerate}
\section{Results and discussion \label{results}}
\subsection{Analysis of difference in learning}
In order to check if there is any difference in learning between students working within the tutor-web and students doing traditional homework model (\ref{anova}) was fitted to the data. The \texttt{lme} function, belonging to the \texttt{nlme} package \citep{pinheiro3r}, in R was used for the estimation. The interaction term between the treatment effect (tutor-web/traditional) and math experience effect was found to be insignificant (p-value = 0.19) and therefore removed from the model. The model was fitted again and the treatment effect was found to be insignificant (p-value = 0.22). After removing the insignificant terms the model becomes
\begin{equation}
y_{ijkl} = \beta_{2j} + \beta_{4k} + b_{l} + \epsilon_{ijkl}
\end{equation}
where $\beta_2$ is the fixed math experience effect ($j = 1,2$), $\beta_4$ is the exam effect ($k = 1,2,3,4$) and $b$ is the random student effect ($l = 1,2,...,157$).

The estimated confidence interval for the difference between the groups doing regular homework and working within the tutor-web was $-0.54 < \beta_1 < 0.12$.
According to this analysis it can not be concluded that there is a difference in mean test scores between students working within the tutor-web and students doing regular homework. This indicates that time and money can be saved by using the tutor-web without reducing the quality of the teaching. 
\subsection{Analysis of quiz question responses}
In order to investigate the difficulty of the items, models (\ref{m1}) - (\ref{m4}) were fitted to the question responses. Four questions banks were used with different number of questions in each bank (Table \ref{questions}). The average number of times the questions were answered was also different between the banks.
Functions from the \texttt{ltm} package in R \citep{rizopoulos2006ltm} were used for the estimation. The models were compared using a likelihood ratio test. 
\begin{table}
\caption{Number of questions and average number of times questions were answered in the item banks.}
\begin{center}
\begin{tabular}{ccc}
\hline
\hline
\\[-5pt]
\multicolumn{1}{c}{Subject covered} & Number of questions & Average number of times \\
&in the bank&questions were answered\\
\hline
Discrete distributions & 70 & 51.38\\
Continuous distributions & 63 & 17.48\\
Hypotheses concerning means & 65 & 49.34\\
Contingency tables & 55 & 43.76\\
\hline
\end{tabular}
\end{center}
\label{questions}
\end{table}

The four question banks were analysed separately. 
The first bank consisted of questions about discrete distributions. The fit did not improve when using the more complex models (\ref{m2})-(\ref{m4}) so it was decided to use model (\ref{m1}) to describe the data. The same was found to be the case for the second question bank containing questions about continuous distributions. 
The third question bank consisted of questions on hypotheses regarding means. In that case model (\ref{m3}) lead to significant better fit than the more simple models and was therefore used to describe the data. For the last question bank, containing questions about contingency tables, the fit was not improved using the more complex models. 

\begin{figure}[t]
\caption{Histogram of the probability of a correct answer for the average student resulting from the model fits.}
\centering\includegraphics[scale = 0.7]{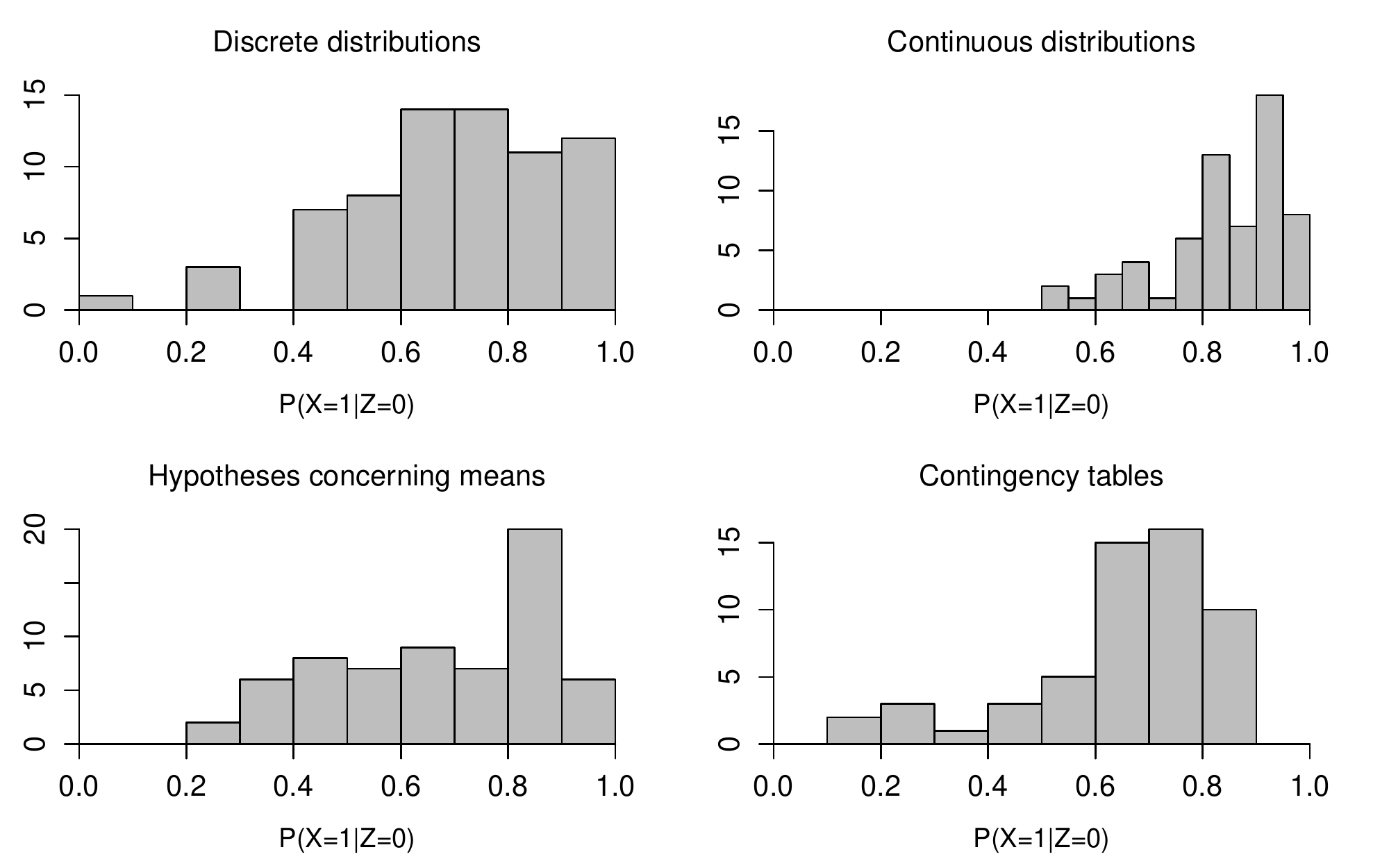}
\label{betahist}
\end{figure}

The models were used to predict the probability of a correct answer to the questions for an average student (Fig \ref{betahist}). 
If the question banks consisted of questions with broad range of difficulty levels and about equal number of easy and difficult questions the distribution would range from 0 to 1 and be symmetric around 0.5. It is clear by looking at the figure that it is not the case for the four question banks.

Looking more closely at the figure it can be seen that the mass of the distributions is mostly centred to the right for all four question banks. This implies that the number of easy questions is higher than the number of difficult questions in the question pools. The situations is worst in the second pool where the average student has more than 50\% chance of answering each questions correctly. 

These results provide important guidelines how to compose additional questions for the tutor-web. Special attention will be made on including more difficult questions in the question banks in the future. This will become even more important when more advanced item allocation schemes are used (Section \ref{futurework}).

\subsection{Students feedback}
\begin{figure}[t]
\caption{The students experience using the tutor-web.}
\centering\includegraphics[scale = 0.61]{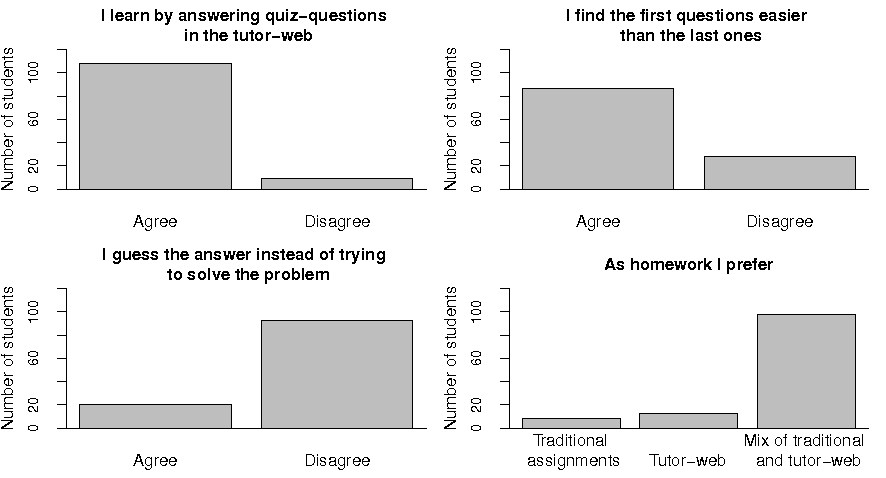}
\label{survey}
\end{figure}

The answers to the four questions regarding the students experience using the tutor-web were tallied (Fig. \ref{survey}).
Some 92\% of the students claimed that they learned by using the tutor-web and about 75\% of the students found the last questions easier than the first ones (implying learning). In the third question, around 17\% of the students said that they guessed the answer instead of trying to solve the problem.
 
The students could choose from three answers in the fourth question where they were asked whether they preferred tutor-web assignments, traditional homework or a mixture of both as homework. 
7\% of the students preferred handing in traditional homework, 11\% the tutor-web while 82\% of the students wanted a mixture of traditional assignments and the tutor-web as homework. 

The students experience using the tutor-web seems in general to be very positive. The students claim that they learn by using the system and many of the students described it as nice alternative to written assignment during the course. It is of some concern that 17\% of the students guess the answer instead of trying to solve the problems. 
It is interesting that 82\% of the students preferred a mixture of the tutor-web and written assignments. During the course, the comments from the students were often that even though the tutor-web is an fun environment to work in it nothing can replace the traditional pencil and paper assignments altogether. 

\section{Future work \label{futurework}}
The tutor-web is an ongoing research project and the system will be developed further in the nearest future. Two aspects of the system particularly need further development, namely educational material content and the item allocation algorithm. In addition to improving the system, new methods to analyse question responses need to be developed. 

\subsection{Educational material content}
One goal of the tutor-web project is to set up a repository of educational material for a BSc degree in mathematics and and a MSc degree in applied statistics. 
Some courses are ready for general use with slides, handouts and questions while other are only placeholders waiting to be filled up with material. 
In addition to university courses, a high school mathematics overview is already available with a total of 390 quiz questions.

\subsection{Item allocation}
When the experiments described here were conducted items were allocated randomly to students taking quizzes. 
An interesting question is: how should items be allocated so the students get the most out of the quiz sessions? In computerized adaptive testing (CAT) \citep{wainer2000computerized} the Point Fisher Information method is often used to allocate items where the most informative item is chosen for the student. The purpose of CAT, however, is to measure student ability while the purpose of the quizzes in the tutor-web is to increase learning. A more appropriate item allocation scheme could be:
\begin{enumerate}
 \item Select easy questions in the beginning and more difficult questions as the grade gets higher.
 \item Select questions the student has answered incorrectly in the past.  
 \item Select questions from old material to refresh the students memory.
 \item Select questions from prerequisite material if the student does not appear to be learning.  
\end{enumerate}

\begin{figure}[t]
\caption{Probability mass function for item allocations with 50 questions and $q = 0.85$.}
\centering\includegraphics[scale = 0.6]{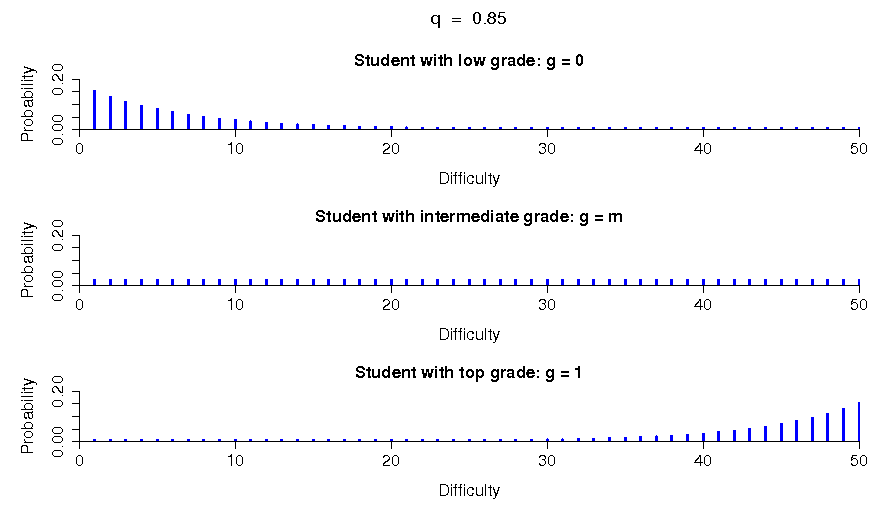}
\label{pmf}
\end{figure}

The first part of the scheme has already been implemented in the tutor-web. Instead of using a uniform probability distribution when allocating items (as done when the experiments were performed) a probability mass function that depends on the grade of the student is used. The idea is that the system collects information on how often a question is allocated and how often it is answered correctly. For this purpose the difficulty of the question is simply calculated as 
$$1 - \frac{\text{number of correct answers}}{\text{number of times question is answered}}.$$ 
The questions are then ranked according to their difficulty, from the easiest question to the most difficult one. A student with a low grade should have higher probability of getting the first questions after the ranking while a student with a high grade should have higher probabilities of getting the last questions. The mass of the probability function should therefore move towards the difficult questions as the grade goes up. This can be done with the following probability mass function (Fig \ref{pmf})
\begin{equation}
p(r) =
\begin{cases}
\displaystyle
\frac{q^r}{\sum_{r=1}^{I}{q^r}} \cdot \frac{m-g}{m} + \frac{g}{I \cdot m} & \text{if } g < m,\\
\\
\displaystyle
\frac{q^{I-r+1}}{\sum_{r=1}^{I}{q^{I-r+1}}}\cdot \frac{g-m}{1-m} + \frac{g-1}{I\cdot(1-m)} & \text{if } g \geq m.
\end{cases}
\end{equation}
where $q$ is a constant ($0 \leq q \leq 1$) controlling the steepness of the function, $I$ is the number of questions, $r$ is the difficulty rank of the question ($r = 1,2,...I$), $g$ is the grade of the student ($0 \leq g \leq 1$) and $m$ is a predefined constant ($0 < m < 1$). When $g = m$ the probability mass function is uniform. When $g < m$ the mass is mostly located at the easy questions while the mass is mostly located at the difficult questions when $g > m$.

\subsection{Analysis of quiz question responses}
Some similarities are between students taking quizzes in the tutor-web and students taking a computerized adaptive test. There is a fundamental difference though since the main purpose of the quizzes in the tutor-web are normally not to measure ability but to enhance learning. Item response theory type models which are often used in CAT do not take learning into account. New models should therefore be developed that take learning in a dynamic environment into account. 
Also, restricting consideration to a small set of models with a fixed parametrization can be dangerous. Goodness-of-fit tests of these models will likely lead to
their rejection by collecting more data. The simple fact that having two students (A,B) and two items (a,b) where student A knows the answer to item a and not b, and vice versa with student B, implies that a general interaction model is needed to describe the data.

\section{Conclusions \label{conclusion}}
An educational system, the tutor-web has been developed and used for educational research. Educational material in mathematics and statistics can be found in the system which is open to everybody having access to the Web. In addition to well structured educational material the system offers quizzes with the purpose of increase learning rather than evaluation. 

A randomized cross-over experiment was conducted to asses the difference in learning between students using the tutor-web and students doing regular homework. The difference in mean exam scores between the groups was not significant. This indicates that time and money can be saved by using the tutor-web as homework in stead of written assignments to some extent. The students' experience using the tutor-web was very positive but the students preferred a mixture of tutor-web and written assignments as homework. When analysing the question banks used it was found that more difficult questions need to be added to the item banks.

The system will be developed further with the main focus being on educational content and item allocation. 

\section{Acknowledgement \label{thanks}}
The tutor-web project has been supported by the Marine Research Institute of Iceland, the United Nations University and the University of Iceland.      

\newpage
\bibliography{biblioarxiv}{}

\begin{thebibliography}{11}
\newcommand{\enquote}[1]{``#1''}
\expandafter\ifx\csname natexlab\endcsname\relax\def\natexlab#1{#1}\fi

\bibitem[{Goossens et~al.(1994)Goossens, Mittelbach, and
  Samarin}]{goossens1994latex}
Goossens, M., Mittelbach, F., and Samarin, A. (1994), \textit{The LATEX
  companion}, Citeseer.

\bibitem[{Latteier et~al.(2002)Latteier, Pelletier, McDonough, and
  Sabaini}]{latteier2002zope}
Latteier, A., Pelletier, M., McDonough, C., and Sabaini, P. (2002), \textit{The
  Zope Book}, New Riders.

\bibitem[{Lord(1980)}]{lord1980applications}
Lord, F. (1980), \textit{Applications of item response theory to practical
  testing problems}, L. Erlbaum Associates Hillsdale, NJ.

\bibitem[{Nagle(2010)}]{nagle2010plone}
Nagle, R. (2010), \textit{A User's guide to Plone 4}, Enfold Systems Inc.

\bibitem[{Pinheiro et~al.(2011)Pinheiro, Bates, DebRoy, Sarkar, and {R
  Development Core Team}}]{pinheiro3r}
Pinheiro, J., Bates, D., DebRoy, S., Sarkar, D., and {R Development Core Team}
  (2011), \textit{nlme: Linear and Nonlinear Mixed Effects Models}, r package
  version 3.1-102.

\bibitem[{{R Development Core Team}(2011)}]{R2011}
{R Development Core Team} (2011), \textit{R: A Language and Environment for
  Statistical Computing}, R Foundation for Statistical Computing, Vienna,
  Austria, {ISBN} 3-900051-07-0.

\bibitem[{Rizopoulos(2006)}]{rizopoulos2006ltm}
Rizopoulos, D. (2006), \enquote{ltm: An R package for latent variable modeling
  and item response theory analyses,} \textit{Journal of Statistical Software},
  17, 1--25.

\bibitem[{Stefansson(2004)}]{stefansson2004tutor}
Stefansson, G. (2004), \enquote{The tutor-web: An educational system for
  classroom presentation, evaluation and self-study,} \textit{Computers \&
  Education}, 43, 315--343.

\bibitem[{Tantau et~al.(2011)Tantau, Wright, and Miletić}]{number2011beamer}
Tantau, T., Wright, J., and Miletić, V. (2011), \enquote{The beamer class,} .

\bibitem[{Wainer(2000)}]{wainer2000computerized}
Wainer, H. (2000), \textit{Computerized adaptive testing}, L. Erlbaum
  Associates Hillsdale, NJ.

\bibitem[{Williams and Kelley(2010)}]{williamsgnuplot}
Williams, T. and Kelley, C. (2010), \enquote{Gnuplot 4.4-An Interactive
  Plotting Program,} .

\end{thebibliography}
\bibliographystyle{asa.bst}

\end{document}